\documentclass[aps,prl,preprint,superscriptaddress]{revtex4-2}

\usepackage{graphicx}% Include figure files
\usepackage{dcolumn}% Align table columns on decimal point
\usepackage{bm}% bold math
\usepackage{mhchem}
\usepackage{siunitx}

\begin{document}

\title{Intrinsic Maximum Light Absorption in Laser-Field-Driven Growth of Highly Ordered Silicon Nanowire Arrays}
% \thanks{A footnote to the article title} 

% 作者1
\author{Jin Qin}

% 作者2 (通讯作者)
\author{Zhikun Liu}
\email{liuzhikun@sjtu.edu.cn} 

\affiliation{
    School of Mechanical Engineering, Shanghai Jiao Tong University, Shanghai 200240, China
}

\date{\today}% It is always \today, today,

\begin{abstract}
We provide direct experimental evidence for a state-selection principle in a far-from-equilibrium system. Using the laser-driven growth of silicon nanowires as a uniquely clean and quantifiable platform, we show that a long-range ordered array emerges as the system spontaneously selects the periodicity that maximizes its collective light absorption. This establishes a direct, measurable link between a maximum dissipation/absorption principle and emergent structural order.  Our results thus offer a concrete test for models of non-equilibrium self-organization.

\end{abstract}

\maketitle

%\tableofcontents

In non-equilibrium systems, individual units behave collectively and cooperatively, self-assembling into diverse spatial patterns to adapt to their environment at the cost of energy dissipation\cite{Cross1993}\cite{Feinerman2018}\cite{Priimagi2024}\cite{Xu2024}, offering transformative opportunities in intelligent materials, maskless nanofabrication and the understanding of life's foundational dynamics\cite{Ornes2017}\cite{Fang2019}. However, due to the complexity of these nonlinear processes, it remains unclear how systems select their final structures, posing significant challenges in predicting or controlling the resulting patterns. It has been proposed that, within the vast space of possible structures, configurations that are capable of absorbing more energy from their driving source are statistically more likely to emerge as the final steady-state configurations\cite{Jarzynski1997}\cite{Dewar2005}\cite{England2015}. In light-driven systems, for example, interference effects significantly influence the energy absorption for different shapes, leading to the selection of specific configurations that better respond to light source\cite{Ito2013}\cite{Ghalawat2024}. Over the past decade, simulations and experiments across various systems—such as periodic force-driven chemical reactions (LIPSS)\cite{Kachman2017}, voltage-driven assembly of conducting beads\cite{Kondepudi2015}, light-driven thermo-optical resonator arrays\cite{Ropp2018}, and laser-induced periodic surface structures\cite{Zou2020}\cite{Bonse2020} have demonstrated that the steady-state configurations of dissipative structures can indeed dissipate more energy than their initial arrangements.
Despite these advances, precisely correlating structural selection with energy dissipation remains challenging. Irregularity and nonuniformity in non-equilibrium structures obscure direct links between structural features and energy absorption, making it difficult to establish clear selection rules. Additionally, the interplay of multiple forces—including heat, light, and chemical gradients—often complicates the identification of the dominant driving mechanism. Finally, the vast structural space of nonequilibrium systems lacks an internal parameter to guide exploration of all relevant configurations.

In this study, we address these challenges by introducing a novel system that enables the near-unconstrained evolution of highly regular and uniform planar silicon nanowire arrays (SiNWAs) utilizing femtosecond laser-driven chemical vapor deposition [Fig. 1(a)], where under energy-limited conditions, the silicon's evolution must be exceptionally precise to achieve efficient energy absorption and sustain growth through strong resonance with the laser. This approach yields exceptional structural order, enabling a direct and quantitative investigation of light absorption's role in directing self-organization. The use of ultrafast laser pulses minimizes thermal impact during energy deposition, while employing a transparent substrate ensures that laser energy is selectively absorbed by the growing silicon nanostructures, isolating the growth process from substrate-induced thermal effects. Consequently, the silicon growth is predominantly driven by the light it absorbs, allowing it to evolve towards its intrinsic optimal structure with minimal external perturbations. The resulting SiNWAs exhibit remarkable periodicity and long-range order, simplifying the structural space to a single dominant parameter – the array period. This simplicity allows for a direct and quantitative correlation between energy dissipation and structural selection, offering an unprecedented platform for studying self-organization in non-equilibrium systems.

SiNWAs were synthesized using a femtosecond laser (wavelength: 515 nm, pulse duration: 250 fs, repetition rate: 500 kHz) in a controlled vacuum environment with silane (\ce{SiH4}) as the silicon precursor. Growth occurred in a laser-limited regime, with the peak laser intensity set at \SI{5e9}{\watt\per\centi\meter\squared} to \SI{1.3e10}{\watt\per\centi\meter\squared},  two orders of magnitude lower than those used in previous methods\cite{He2016}. On transparent substrates such as \ce{Al2O3}, MgO, and \ce{SiO2}, direct silane decomposition was not feasible, requiring a catalyst to trigger initial nucleation.

\begin{figure}
    \centering
    \includegraphics[width=0.9\textwidth]{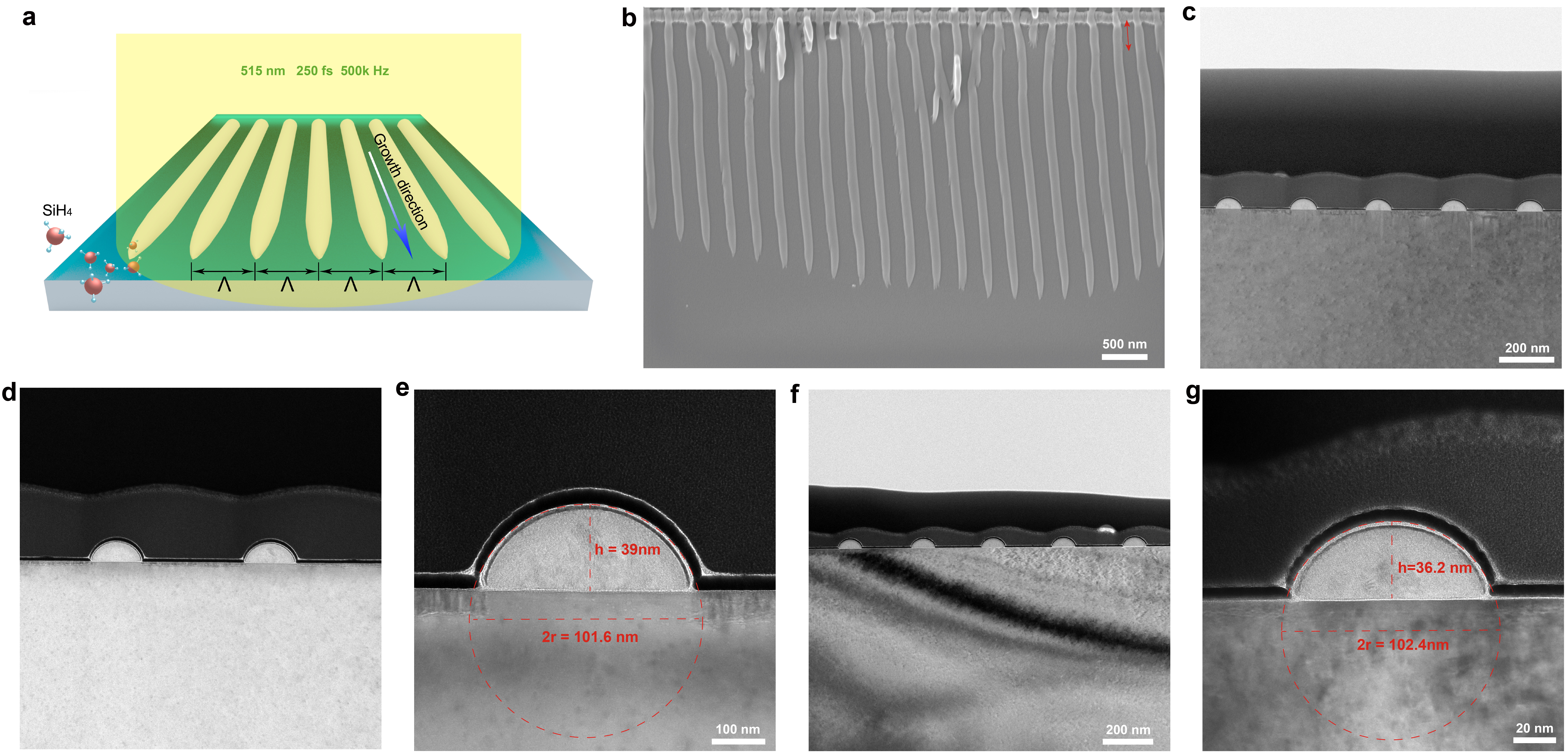}
    \caption{Laser driven growth of Silicon nanowire array. (a) Schematic illustration of the spontaneous growth of Si nanowires driven by laser irradiation. (b) Typical SEM image of a SiNWA grown on an \ce{Al2O3} substrate, induced by a long silver wire. The red arrow indicates the polarization direction of incident laser. (c-e) Cross sectional TEM images of a SiNWA with a period of 274nm grown on \ce{Al2O3} substrate at a laser peak intensity of \SI{5.4e9}{\watt\per\centi\meter\squared}, with a \ce{SiH4} flow rate of 30 sccm and an \ce{H2}  flow rate of 100 sccm. (f, g) SiNWA with a period of 279 nm, grown under the same conditions but with a reduced silane flow rate of 6 sccm.}
    \label{fig:enter-label}
\end{figure}
Silver particles and wires served as catalysts to facilitate silicon growth. We observed that using silver wires resulted in the formation of a highly regular planar SiNWA, as illustrated in Fig. 1(b). The SiNWA comprises equidistantly aligned nanowires oriented parallel to the laser polarization direction. Precise dimensions of the SiNWA were determined using FIB and TEM analyses. Figure. 1(c) presents a cross-sectional TEM image of a nanowire array grown under 30 sccm \ce{SiH4} and 100 sccm \ce{H2} at a peak intensity of \SI{5.4e9}{\watt\per\centi\meter\squared}. The resulting nanowires exhibited a periodicity of 274 nm. Cross-sectional TEM images of representative nanowires, presented in Figs. 1(d) and (e), reveal structures resembling segments of minor circles, with a height of 39 nm and a radius of curvature of 50.8 nm. The nanowires within the array displayed a narrowsize distribution, with heights of different nanowires ranging from 39 ± 0.4nm—a level of uniformity significantly superior to that achieved by conventional methods\cite{McIntyre2020}\cite{Wendisch2020}.

The nanowire cross-sectional shape and period depend on both growth conditions and substrate material. For example, also on \ce{Al2O3} substrate, an array grown with 6 sccm \ce{SiH4} and 100 sccm \ce{H2} under \SI{5.4e9}{\watt\per\centi\meter\squared} laser power yields nanowires with a height of 36.2 nm, a radius of 51.2 nm, and a period of 279 nm, as shown in Fig.1 (f) and (g). The height distribution of nanowires was 36.2±0.6nm. Using \ce{SiO2} as the substrate under same gas conditions and peak intensity of \SI{6.1e9}{\watt\per\centi\meter\squared} results in nanowires with a height of 40.3 nm, a radius of 48.3 nm, and a period of 306 nm.  We observe a trade-off at higher laser intensities, where the nanowire crystallinity improves at the cost of increased size fluctuation [FIG. S6]. We attribute this to a competition between two shaping forces at the growth front: the isotropic surface tension, which favors a uniform circular cross-section, and the anisotropic crystallographic faceting, which is enhanced at higher laser intensity and leads to growth instabilities.

 SiNWA growth occurred in a laser-limited environment, where silicon growth exhibited strong energy-seeking behavior, as demonstrated later. The laser-limited regime was critical for ordered SiNWA growth; in this regime, a silver catalyst was needed, and the length of the silver wire defined the boundary of the SiNWA. At laser intensities exceeding \SI{3.5e10}{\watt\per\centi\meter\squared} silicon growth occurred directly on the \ce{Al2O3} substrate, even without metal catalysts. This led to irregular and uncontrollable silicon structures, which are beyond the scope of this study.

\begin{figure}
    \centering
    \includegraphics[width=0.9\textwidth]{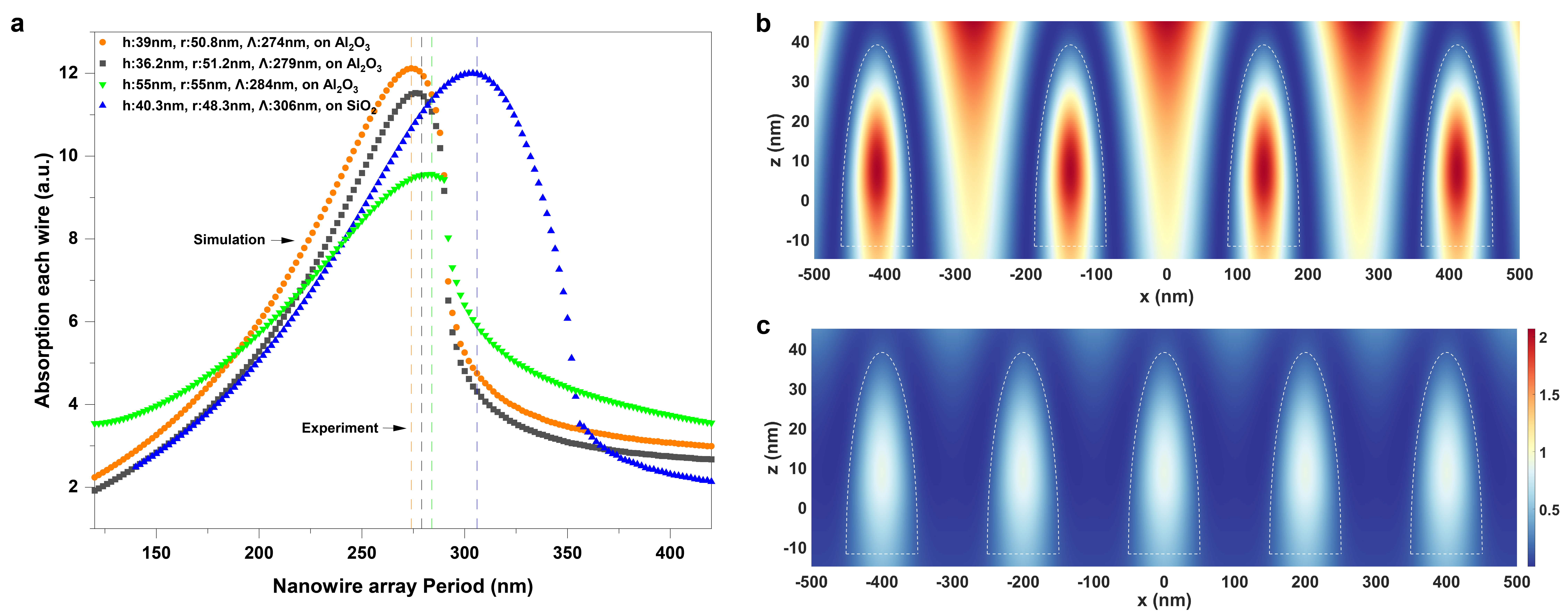}
    \caption{Silicon nanowires maximize light absorption by selecting a specific array periodicity. (a) Calculated absorption per nanowire versus array period, using the experimentally measured nanowire cross-section as the repeating unit. The absorption peak aligns precisely with the experimentally observed period (vertical dashed line). (b),(c) Cross-sectional electric field intensity distribution, \( (E/E_0)^2 \), simulated for the nanowire array (nanowire unit cell: \textit{r}=50.8~nm, \textit{h}=39~nm). The field is strongly enhanced inside the nanowire at (b) the optimal period (274~nm), corresponding to the absorption peak, but is significantly weaker at (c) a non-optimal period (200~nm).}
    \label{fig:enter-label}
\end{figure}

To investigate why SiNWAs preferentially select specific periodicities, we simulated how changes in periodicity influence the light absorption of each nanowire, considering that the growth-driving force is solely derived from the laser energy absorbed. Using the individual nanowire depicted in Fig. 1(e) as the repeating unit of an infinitely large array, we plotted the absorptivity per nanowire as a function of the array period, as shown by the orange curve in Fig. 2(a). Due to strong interference, the array period significantly determines the field distribution inside the nanowire and its light absorption capability. The maximum calculated absorptivity of 12.12 occurs at a period of 274 nm, perfectly matching the experimentally measured period of 274 nm. This indicates that the system selects a period that maximizes light absorption per nanowire, thereby enhancing the growth rate and entropy production rate.

For comparison, the absorptivity per nanowire at the resonant state (12.12) was three times greater than that of an array with a very large period of \SI{2}{\micro\meter} (3.29), where nanowires can be approximated as non-interacting, isolated structures.

The periods of SiNWAs were related to the nanowire cross-sectional shape and the refractive index of the substrate. Investigations of various SiNWAs show that the experimentally measured period consistently closely matches the calculated light absorption peak. Since the driving force of the growth was the light absorbed by the preexisting array, this indicates that stable growth occurred in a resonant state, where the energy dissipation rate and silicon growth rate were at their highest.

Single silicon nanowires, the fundamental building blocks of SiNWA, can be grown from catalyst particles smaller than 200 nm. We demonstrated that the spontaneous growth of silicon into one-dimensional nanowires with sharp tips occurs because this structure achieves an exceptional effectiveness to capture laser energy and promote the growth of itself under low-intensity laser conditions. The silicon growth process proceeds in two distinct stages: initial nucleation on silver nanoparticles, followed by self-sustained elongation. In the first stage, silver nanoparticles act as catalysts that strongly absorb 515 nm laser light due to plasmonic effects. At relatively low laser intensities, silver particles become gradually encapsulated by silicon due to its high reactivity [Fig. 3(a)]. However, because the plasmonic field of silver nanoparticles is short-ranged, extending only a few nanometers\cite{Yang2020}, continued silicon growth requires direct energy absorption by silicon itself. This second stage, where the growth becomes intrinsically dependent on the silicon nanostructure, is therefore the regime of interest for the self-organization phenomena we describe. We find that silicon extends beyond the silver catalysts and preferentially grows along the laser polarization direction [Fig. 3(b)] only when the peak laser intensity exceeds \SI{4.8e9}{\watt\per\centi\meter\squared}. Notably, the system consistently selects nanowires with sharp tips as the final product [Fig.3(c) and (f)], achieving a 100\% growth success rate for all catalysts. While nanowires predominantly align with the laser polarization direction, minor random bending is occasionally observed. Unlike the vapor-liquid-solid (VLS) mechanism\cite{Rathi2013}, here, silver nanoparticles contribute solely to the initial nucleation of silicon and play no role in subsequent nanowire elongation. Consequently, nanowire dimensions are independent of silver particle size and instead exhibit self-regulated growth. Apart from the tip region, each nanowire maintains a consistent cross-sectional profile along its entire length.
\begin{figure}
    \centering
    \includegraphics[width=0.5\textwidth]{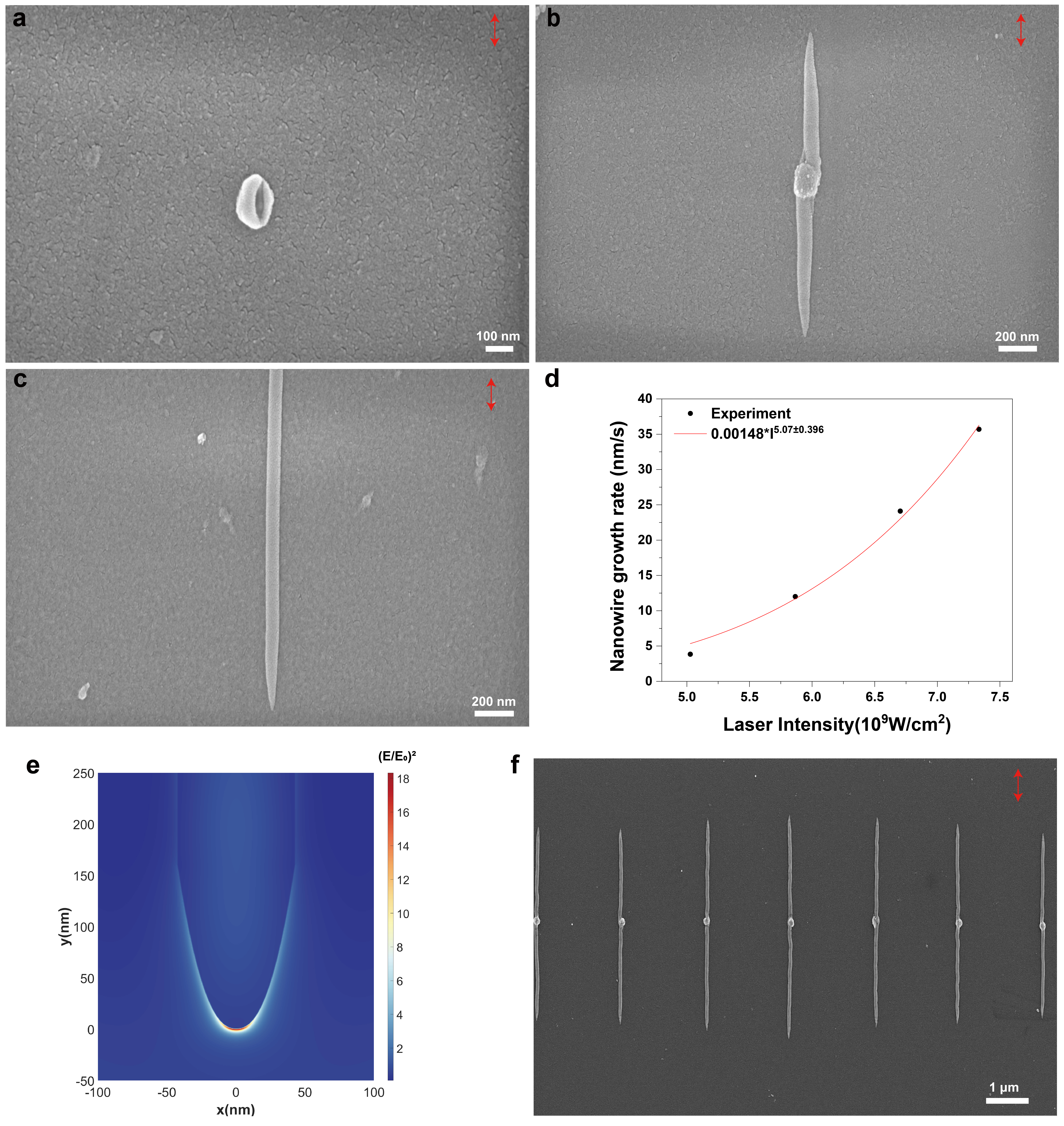}
    \caption{Silicon nanowires growth driven by  tip-field enhancement and multiphoton absorption. (a) SEM image of the initial growth stage, where a Ag  nanoparticle is encapsulated by silicon. (b) A silicon nanowire emerges from the Ag particle and propagates along the \ce{Al2O3} substrate. (c) A fully grown Silicon nanowire exhibiting a uniform diameter and a sharp, well-defined tip. (d) The dependence of the nanowire growth rate on the peak intensity of the incident laser.(e) Simulated electric field intensity distribution at the SiNW tip, 2nm above the substrate, for an incident laser polarized parallel to the nanowire length. (f) Demonstration of controlled, byproduct-free growth of isolated silicon nanowires. }
    \label{fig:enter-label}
\end{figure}

The nanowire growth is driven by a highly nonlinear, non-thermal process. This is evidenced by the growth rate's strong dependence on laser intensity, scaling as \( I^5\) [Fig. 3(d)], which is characteristic of a five-photon absorption process. This observation aligns with previous reports indicating that a minimum of 8–11 eV is required to initiate various steps essential for silicon growth\cite{Berkowitz1987}\cite{Fons1994}, as the photons used here (515 nm) have an energy of only 2.4 eV. Contributions from thermal effects are ruled out by the growth dependence on peak power over pulse energy\cite{Chichkov1996-jf} and a minimal calculated temperature increase(80\text{K}).

Characterization by TEM and AFM reveals that the nanowires possess highly tapered tips where the height gradually approaches zero at the apex. When nanowires with such tip structures were exposed to normally incident laser light, the electric field at their tips was significantly enhanced to \( 18.3 E_0^2 \) [Fig. 3(e)] due to the light-rod effect\cite{Tian2022}. Because the growth rate is highly dependent on the laser field strength, this enhancement resulted in the growth rates that were six orders of magnitude faster than those of other structures lacking similar field enhancement. As a result, nanowires were the only products observed[Fig. 3(f)]. Furthermore, previously deposited nanowires did not thicken during extended laser exposure due to the low growth rate at their surfaces. This resulted in a consistent cross-sectional shape along the entire length.

Silicon spontaneously grow in to optimal silicon nanowire array via two strategy, first, it grow into a lager array through cooporation effect, second, the arrays adjust the period toward an optimal state. First we show individual silicon nanowires exhibit a strong tendency to align and collectively grow into an array. For a given silver wire, the system self-assembled into an array contains maximum number of nanowires possible, constrained only by the length of the silver wire and the inter-nanowire spacing[Fig. 4(a)]. To understand why nanowires spontaneously form organized arrays, we calculated the changes in the laser field when one or more nanowires were irradiated. When SiNWA was exposed to normal incident laser, the curved surfaces of the wires allowed part of the incident laser light to propagate parallel to the substrate, causing inter-nanowire interactions and modifying the laser field distribution. Consequently, while the electric field strength at the tip of an isolated nanowire reached \( 18.3 E_0^2 \) (where \( E_0 \) is the incident electric field strength), this increased to \( 19.7 E_0^2 \) and \( 24 E_0^2 \) when two or three parallel nanowires, spaced 268 nm apart, were irradiated.

This field enhancement extended along the entire nanowire rather than being confined to the tip. Because light-induced enhancement is sensitive to the nanowire tip structure, the increased field across the nanowire provided a more generalized effect. As shown in Figs. 4(b-d), the electric field distribution across the cross-section of an infinitely long, isolated nanowire showed a maximum field intensity of \( 0.7 E_0^2 \). By contrast, for two nanowires spaced 268 nm apart, the maximum field intensity rose to \( 1.1 E_0^2 \), and for three nanowires, it further increased to\( 1.23 E_0^2 \). The cooperative interaction within the array is further evidenced by a decreased growth threshold and enhanced growth rate of the array compared to isolated nanowires. We found the growth threshold for a single nanowire to be \SI{4.8e9}{\watt\per\centi\meter\squared}, whereas a large SiNWA requires only \SI{3.7e9}{\watt\per\centi\meter\squared}. Both under a laser intensity of \SI{5.4e9}{\watt\per\centi\meter\squared}, silane flow of \SI{30}{sccm}, and \ce{H2} flow of \SI{100}{sccm}, the experimentally measured linear growth rates for isolated nanowires and nanowire arrays were \SI{3.8}{\nano\meter\per\second} and \SI{28.3}{\nano\meter\per\second}, respectively. Considering the difference in cross-sectional shape, these correspond to volumetric growth rates of \SI{7311}{\nano\meter\cubed\per\second} and \SI{81108}{\nano\meter\cubed\per\second}. This significant enhancement suggests that the growth driving force within an array is largely nonlocal, primarily contributed by collective cooperative effects.

\begin{figure}
        \centering
        \includegraphics[width=0.9\textwidth]{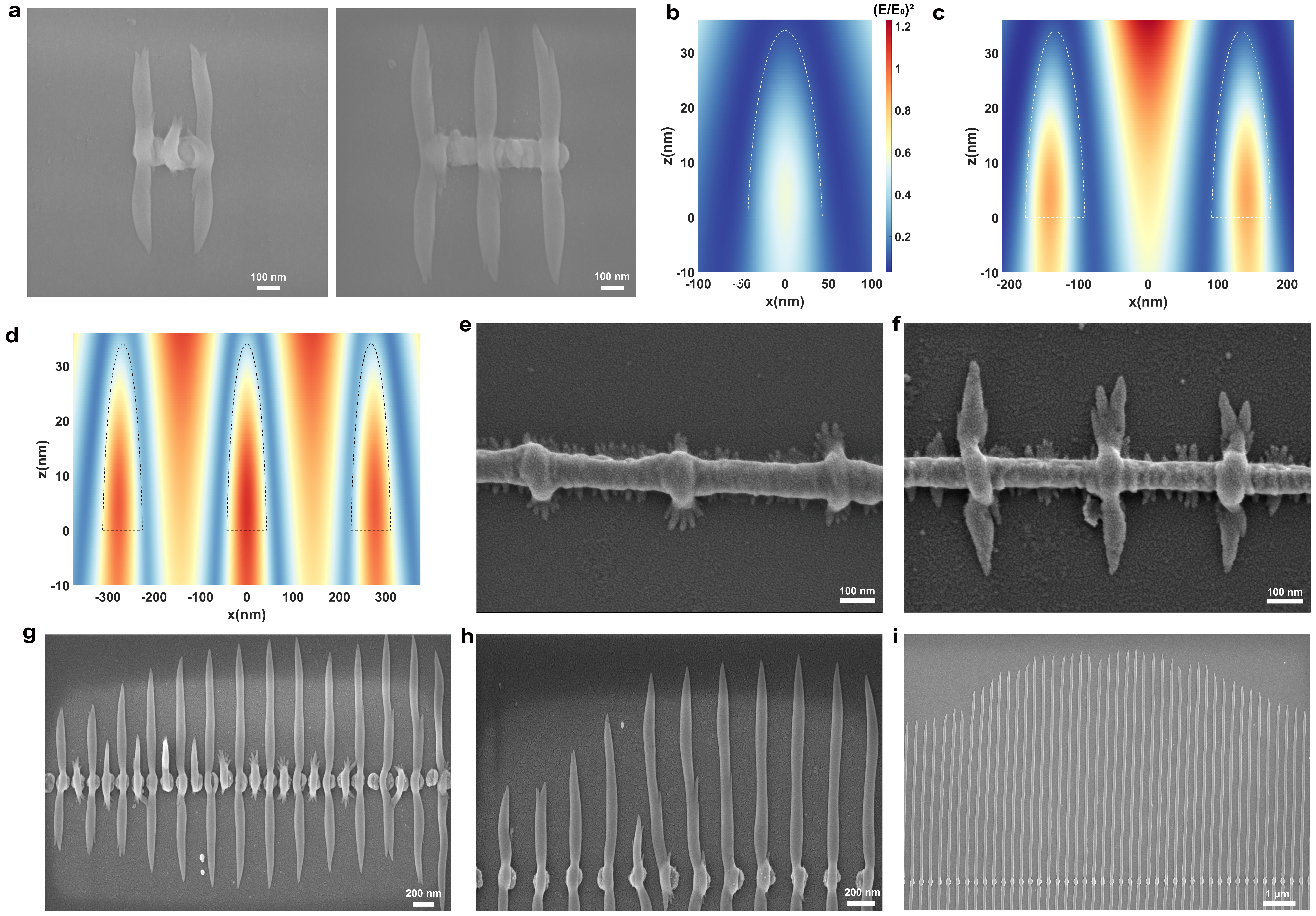}
        \caption{Spontaneous optimal array formation guided by maximum energy absorption. (a) Two and three silicon nanowires arrays were grown from silver wires with lengths of 300 nm (left) and 580 nm (right), respectively. (b-d) Simulated cross-sectional electric field distributions for one, two, and three silicon nanowires, respectively, under laser irradiation. Dashed lines indicate nanowires positions. (e, f) SEM images of the initial Si growth stages on Ag wires, showing the transition from small Si “beards” to periodic, short nanowires. (g-i) SEM images of SiNW arrays grown from Ag particle arrays with periodicities of (g) 140 nm, (h) 240 nm, and (i) 280 nm.}
        \label{fig:enter-label}
  
\end{figure}
In a uniform laser intensity background, periodic nanowire formation breaks translational symmetry. Observations show that silicon explores various structural configurations, ultimately selecting an optimal array through fluctuations, competition, and cooperation processes. This selection process is evident during the early formation of nanowires on a smooth silver wire prepared via wet chemical methods, providing a uniform background for nanowire deposition. Figure 4(e) shows the early growth stage before array formation, with small, randomly distributed silicon "beards" covering the silver wire at a high density (1 cluster per 20 nm). At this time, nanowires begin to form. As competition progresses, most silicon beards cease growing, while periodic nanowires dominate due to cooperative effects [Fig. 4(f)].

Then we show SiNWA  self-correct their period. If the SiNWA is intentionally arranged with a different period, it cannot sustain steady growth at that period (with a slow growth rate); instead, it rapidly adjusts to the new optimal configuration. This behavior is demonstrated in Figs. 4(g) and (h), where nanowires grown from silver particle arrays with initial periods of 140 nm and 240 nm ultimately self-organize into arrays with a period of approximately 280 nm. In contrast, using a silver particle array with a period matching that of the SiNWA results in the formation of a defect-free SiNWA [Fig. 4(i)].

In summary, we have used the strong-field laser-driven growth of silicon nanowire arrays to provide direct, quantitative evidence for a maximum energy absorption principle governing structural selection far from equilibrium\cite{Martyushev2021}. The inherent regularity and isolation of this system make it a uniquely tractable platform, allowing us to deconstruct the self-organization into a hierarchical process: initial field enhancement at individual nanowire tips drives nonlinear growth, which is then cooperatively amplified as the wires assemble into a long-range ordered array. Crucially, we demonstrate that this collective system spontaneously evolves to select the precise periodicity that maximizes its optical absorption. This establishes a direct, measurable link between emergent structural order and the rate of energy dissipation. These findings provide a benchmark for models of non-equilibrium statistical mechanics and offer a new physical basis for understanding related pattern-formation phenomena, such as the decades-old puzzle of LIPSS.

\begin{acknowledgments}
This work was supported by The Science and Technology Commission of Shanghai Municipality under Grant No.19511132300. We thank the Advanced Electronic Materials and Devices of Shanghai Jiao Tong University for nanofabrication, and Instrumental Analysis Center of Shanghai Jiao Tong University for characterization.
\end{acknowledgments}

\bibliography{apssamp}% Produces the bibliography via BibTeX.

\end{document}